\documentclass[sigconf]{acmart} 
\AtBeginDocument{%
  }

\usepackage{multirow}
\usepackage{graphicx}
\usepackage{xcolor}
\usepackage{colortbl}
\usepackage{rotating}
\usepackage{amsmath}
\usepackage{hyperref}
\usepackage{subcaption}
\usepackage{booktabs}
\usepackage{makecell}

\usepackage{xspace}
\newcommand{\name}{\textsc{WigglyEyes}\xspace}

\usepackage{ifthen}
\newboolean{revising}
\setboolean{revising}{false}
\ifthenelse{\boolean{revising}}
{
    \newcommand{\rv}[1]{\textcolor{blue}{#1}}
} {
    \newcommand{\rv}[1]{    #1}
}

\copyrightyear{2025}
\acmYear{2025}
\setcopyright{cc}
\setcctype{by}
\acmConference[ISWC '25]{Proceedings of the 2025 ACM International Symposium on Wearable Computers}{October 12--16, 2025}{Espoo, Finland}
\acmBooktitle{Proceedings of the 2025 ACM International Symposium on Wearable Computers (ISWC '25), October 12--16, 2025, Espoo, Finland}\acmDOI{10.1145/3715071.3750430}
\acmISBN{979-8-4007-1481-8/2025/10}

\begin{document}

\title{\name: Inferring Eye Movements from Keypress Data}

\author{Yujun Zhu}
\affiliation{%
  \institution{Aalto University}
  \state{Helsinki}
  \country{Finland}
  }
\email{quintus0505@163.com}

\author{Danqing Shi}
\affiliation{%
  \institution{Aalto University}
  \state{Helsinki}
  \country{Finland}}
\email{danqing.shi@aalto.fi}

\author{Hee-Seung Moon}
\affiliation{%
  \institution{Chung-Ang University}
  \state{Seoul}
  \country{Korea}
  }
\email{hsmoon@cau.ac.kr}

\author{Antti Oulasvirta}
\affiliation{%
  \institution{Aalto University}
  \state{Helsinki}
  \country{Finland}
  }
\email{antti.oulasvirta@aalto.fi}


\begin{abstract}

We present a model for inferring where users look during interaction based on keypress data only. Given a key log, it outputs a scanpath that tells, moment-by-moment, how the user had moved eyes while entering those keys. The model can be used as a proxy for human data in cases where collecting real eye tracking data is expensive or impossible. 
Our technical insight is an inference architecture that considers the individual characteristics of the user, inferred as a low-dimensional parameter vector. We present a novel loss function for synchronizing inferred eye movements with the keypresses. Evaluations on touchscreen typing demonstrate accurate gaze inference.



\end{abstract}


\begin{teaserfigure}
    \centering
    \includegraphics[width=0.95\textwidth]{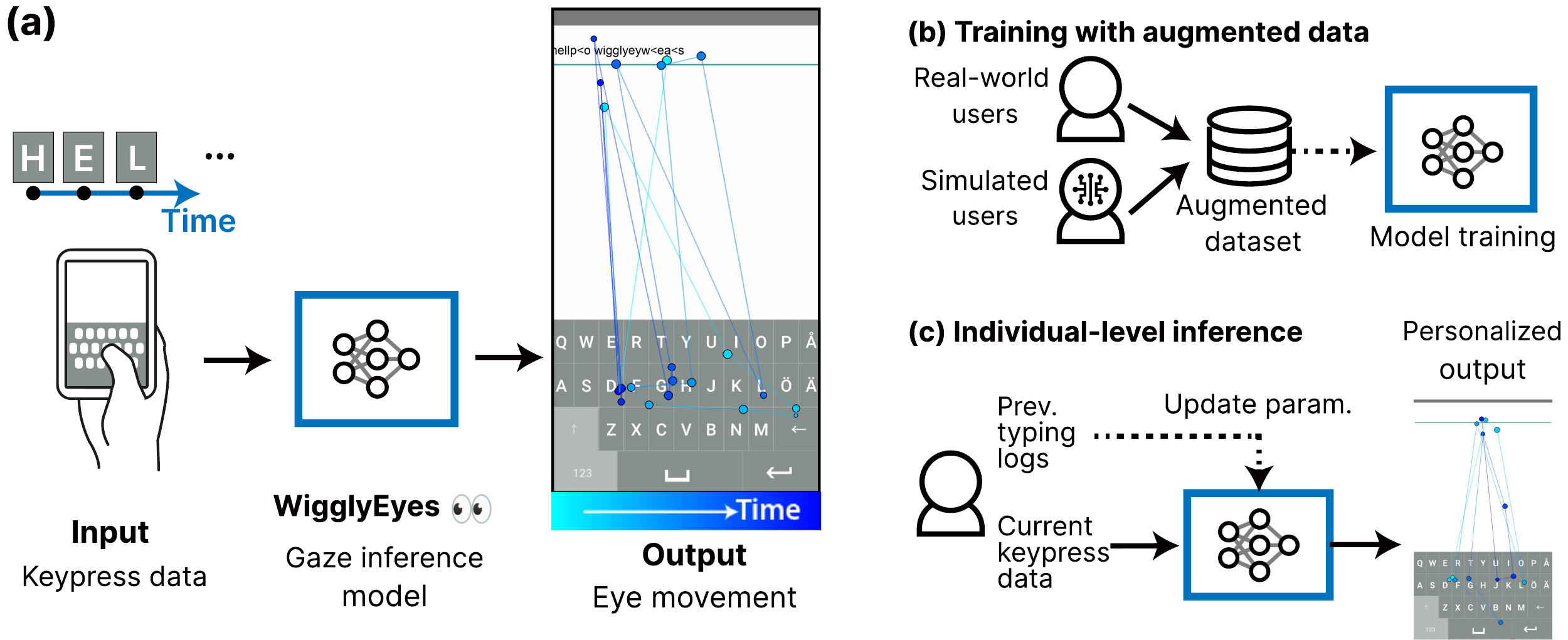}
    \vspace{-4mm}
    \caption{We present \name, a model for gaze inference, designed to infer a user's eye movements based on keypress data. (a) The model takes a sequence of keypresses as input and outputs a sequence of fixations. (b) We train the model using augmented data that combines real-world human data and simulated user data. (c) \name supports individual-level inference by adapting to each user's previous typing log.}
    \label{fig:teaserfig}
\end{teaserfigure}

\maketitle

\section{Introduction}

\begin{figure*}[!t] 
	\centering  
        \includegraphics[width=1.0\textwidth]{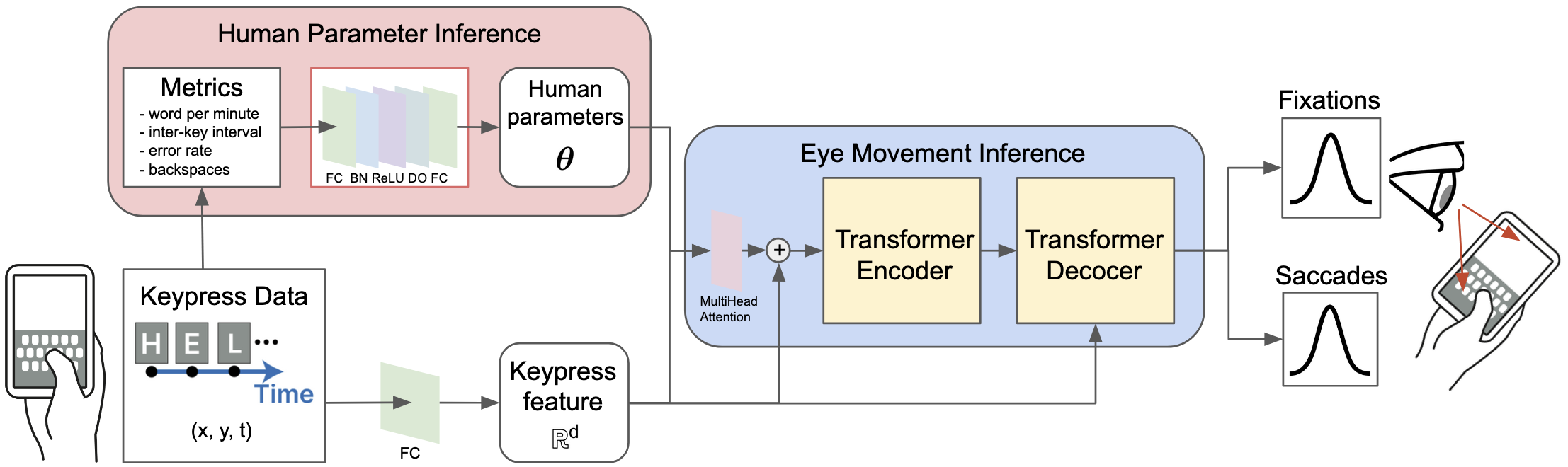}
    \caption{\name architecture mapping keypress sequences $(x, y, t)$ to fixation sequences. The human parameter inference module computes typing performance metrics from the keypress data and infers the human parameters $\theta$. These parameters are combined with encoded keypress features via multi-head attention and processed by a Transformer to predict eye movements.}
	\label{fig:model}
\end{figure*}

Many computer applications have access to keypress data, but few have access to eye tracking data. However, knowing what users look at would be generally informative. Visual attention hints at what they are interested in, where they make errors, what they struggle with, and generally how they process the task~\cite{holmqvist2011eye, rayner1998eye, jacob2003eye}. 
We propose a computational model called \name for inferring gaze movement from keypress data\footnote{\url{https://github.com/quintus0505/wigglyeyes}}. %
We assume a keylog data that includes the positions and timestamps of touch events on a mobile device. 
The output is a \emph{scanpath}: a sequence of fixations with positions and durations. 
Our key contribution is a real-time deep learning architecture that infers eye movements from keypress data and a novel loss function that models eye-finger coordination during both guidance and validation phases.
%
%

\section{Related Work}

There are numerous models that predict users' deployment of attention on graphical displays either in free-viewing or search tasks. These models can be divided into two classes on account of two of the modelled control processes: bottom-up and top-down~\cite{itti2000saliency}. Bottom-up attention is driven by visually distinct (salient) stimuli~\cite{kummerer2022deepgaze}, such as information that is highlighted on screen~\cite{matzen2017data}. Top-down attention is shaped by user's goal. Visual search is a good example ~\cite{mondal2023gazeformer, shi2025chartist}. Here the task is to predict attention given a display and a target. 
In addition to these two areas, previous work has focused on predicting eye movement from human motions, such as eye prediction from head movements~\cite{nakashima2015saliency, hu2021fixationnet} or body movement~\cite{hu2024eye}. Our study is closely related to the prediction of eye movement from human motions. However, we are exploring a new approach to predicting eye movement from finger movements on a touchscreen for the typing task, which has not been addressed before.
Our study also looks at inference at the level of individuals. 
Unlike traditional methods that require eye movement data, our approach captures personalization using only a short typing log, making it more accessible to applications.

\section{\name: Modeling Approach}

This section introduces the model, \name, which can infer eye movement directly from keypress data. 
Given keypress data $L$, representing a total of $C$ keypresses, we define a sequence of tap events $p_{1:C} = {p_1, \dots, p_C}$, where each tap $p_i = (x_\textit{tap}, y_\textit{tap}, t_\textit{tap})_i$ records the position in pixels $(x_\textit{tap}, y_\textit{tap})$ of the tap and the time $t_\textit{tap}$ between consecutive keypresses.
From this data, we generate a scanpath of length $T$, consisting of a sequence of fixation points $f_{1:T}={f_1, \dots, f_T}$, which capture both spatial and temporal aspects of the user's gaze during typing.
Each fixation point $f_i=(x_\textit{fix}, y_\textit{fix}, t_\textit{fix})_i$ includes the gaze position $(x_\textit{fix}, y_\textit{fix})$, and $t_\textit{fix}$ represents the fixation duration.




\subsection{Model Design}

Our architecture comprises of two main modules (see Figure~\ref{fig:model}). 
The first is the \textit{human parameter inference} module, which infers individual-level latent parameters, referred to as human parameters, that capture key characteristics unique to each user. 
The second module, the \textit{eye movement inference} module, combines the inferred human parameters with keypress data from a given trial to predict the user's eye gaze during that specific trial.




We define \textit{human parameters} here as latent (cognitive) variables that, for example, capture a user's style, preferences, or capabilities.
The process of inferring these human parameters is simulation-based.
Using the touchscreen typing simulation model~\cite{shi2024crtypist, shi2025typoist}, the human parameter inference module learns the relationship between typing performance and human parameters.
Then, the eye movement inference module uses the keypress feature and human parameters to predict fixations. These two features are combined using a multihead attention layer, with the human parameters as the query and the keypress feature as the key and value. 
A Transformer-based encoder and decoder are used to transform the keypress-related input into eye movement data.


\subsection{Loss Function}

We propose a novel loss function for model training that encourages the inferred eye movement to be more human-like spatially and temporally, and also to reproduce select phenomena of eye-hand coordination.
Specifically, the overall loss function $\mathcal{L}$ is a sum of four key loss terms:
$
\mathcal{L} = \mathcal{L}_{\text{sim}} + \mathcal{L}_{\text{len}} + \mathcal{L}_{\text{f}} + \mathcal{L}_{\text{v}}
$
where $\mathcal{L}_{\text{sim}}$ is the fixation similarity loss, $\mathcal{L}_{\text{len}}$ is the scanpath length loss, $\mathcal{L}_{\text{f}}$ is the finger guidance loss, and $\mathcal{L}_{\text{v}}$ is the proofreading loss. We introduce each one of them as follows.

Fixation similarity loss $\mathcal{L}_{\text{sim}}$ ensures that the predicted fixations closely match the ground truth human data in positions and durations. It is a combination of two terms: the mean squared error (MSE) loss and the Multi-match similarity loss term, which is hand-coded based on the multi-match similarity and captures the spatial-temporal similarity between the predicted and true gaze paths \cite{dewhurst2012depends};
Scanpath length loss $\mathcal{L}_{\text{len}}$ is designed to make the model more likely to predict a similar number of fixations as humans, which is defined as:
$\mathcal{L}_{\text{len}} = \frac{1}{N} \sum_{i=1}^{N} \text{BCE}(s_i, \hat{s}_i)$,
where \text{BCE} is the binary cross-entropy loss, $N$ is the number of maximum length of possible gaze points in the sequence, $s_i$ is the predicted padding logits and the true padding labels $\hat{s}_i$;
Finger guidance loss $\mathcal{L}_{\text{f}}$ with two components,
where the first term minimizes spatial discrepancy in visual guidance, while the second measures the discrepancy between the counts of the fixations falling in the window of predicted gaze movement and human gaze movement;
Visual validation loss $\mathcal{L}_{\text{v}}$ encourages predicted gaze to mimic human behavior of switching between keyboard and text areas using two components,
where the first aligns proofreading duration, while the second measures the discrepancy between the number of predicted gaze fixations on the text area and the number of gaze fixations on the text area.

\begin{figure}[!t] 
	\centering  
        \includegraphics[width=0.99\columnwidth]{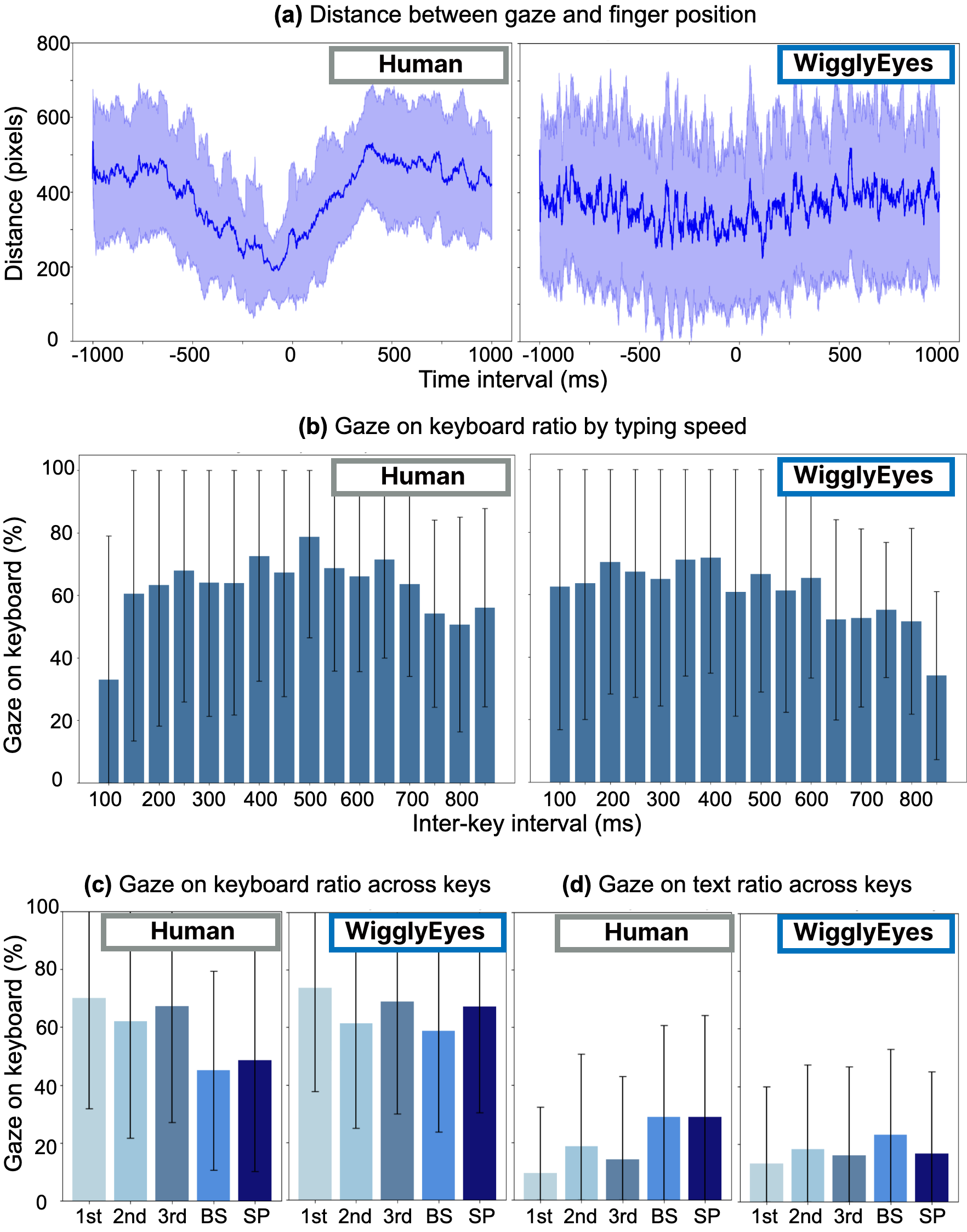}
	\caption{Comparisons between human data and \name. (a) Distance between gaze and finger positions by tapping time; (b) Gaze on keyboard ratio vs. Typing speed; Gaze attention distributions on keyboard (c) and text area (d) across keys, including 1st, 2nd, 3rd row keys, and backspace (BS) and space bar (SP). 
    }
	\label{fig:results}
\end{figure}

\section{Evaluation in Touchscreen Typing}

We evaluated \name in touchscreen typing. 
The gaze movement and typing logs used in our experiments were obtained from \emph{How We Type}~\cite{how_we_type}. It recorded detailed finger and eye-tracking data from 30 participants as they transcribed 20 sentences each, chosen from a set of 75 sentences. The data was collected while the participants were typing on a touchscreen using a customized keyboard. To ensure consistency, the eye tracking and typing logs, captured using different devices, were both resized to a resolution of $1920 \times 1080$ pixels. We randomly choose 25 users and their 285 trails for training, and the other 5 users and their 69 trails for evaluation.

\paragraph{Training}
\label{sec:training_details}
\rv{We first randomly set the three human parameters in CRTypist~\cite{shi2024crtypist}, simulate enough trials and compute the four metrics, creating pairs of human parameters and metrics, and then train the Eye Movement Inference model.} 
The training dataset for \name is the augmented dataset combined with human data and simulation data.
Specifically, we simulated 300 trials using the optimal parameter set reported in CRTypist, based on which the agent can simulate the most human-like typing behavior~\cite{chandramouli2024workflow}. 

\paragraph{Results}

Regarding the average distance between the gaze position predicted by \name and the tapped position, we observe similar behavior. As shown in Figure \ref{fig:results}-a, the distance between the gaze position and the touch point decreases and reaches the lowest point at around 250 ms before the tapping, which is similar to the human data.
For the gaze movement predicted by \name, a similar trend is observed with an initial increase followed by a decrease as shown in Figure \ref{fig:results}-b compared with human data, except for not fully capturing the lower gaze-on-keyboard ratio for IKIs below 100 ms, which may be attributed to the lack of sufficient training data when IKI is low.
\name accurately predicts the gaze-on-keyboard distribution across key rows in Figure \ref{fig:results}-c and d, capturing the human tendency to fixate more on first and third row keys. While the model reproduces most patterns well, it shows minor discrepancies for special keys, particularly underestimating the gaze-on-text ratio for `Space' and `Backspace' keys compared to human data.

\section{Discussion}

While the current framework demonstrates promising results in inferring gaze movements from typing logs, several opportunities exist to enhance its applicability to broader scenarios and domains.
First, the demographic composition of the training dataset, which primarily consists of young Finnish adults, presents an opportunity to explore the model's adaptability to diverse populations. Cultural and behavioral differences in typing patterns may influence gaze dynamics, and expanding the dataset to include participants from varied demographics would strengthen the model's robustness. By incorporating data from users with different linguistic backgrounds, age groups, and typing habits, the model could be refined to capture a wider spectrum of gaze behaviors. Additionally, increasing the dataset size would enable more robust analyses of individual differences, supporting the development of user-specific adaptations and improving overall generalizability.
Second, the model's reliance on correlations between keypresses and gaze patterns highlights the importance of exploring cross-domain adaptations. While the current framework has not been benchmarked against other gaze-inference methods due to the lack of directly comparable baselines in the typing domain, future research could explore cross-domain adaptations of \name
, enabling systematic comparisons with image-based models. Investigating hybrid approaches that combine keypress data with auxiliary signals could improve accuracy.

\begin{acks}
YZ, DS, and AO were supported by the Research Council of Finland (FCAI: 328400, 345604, 341763; Subjective Functions 357578), the ERC (AdG project Artificial User: 101141916.), and Google Grant (DeepTypist).
\end{acks}

\bibliographystyle{ACM-Reference-Format}
\bibliography{wigglyeyes.bib}


\end{document}